\documentclass{article}
\usepackage{appendix}
\usepackage{adjustbox}
\usepackage{subcaption}
\usepackage{graphicx}
\usepackage{multirow}
\usepackage{booktabs}
\usepackage{algorithm}
\usepackage{algpseudocode}
\usepackage{amsmath}

\usepackage[a4paper,
            bindingoffset=0.2in,
            left=1in,
            right=1in,
            top=1in,
            bottom=1in,
            footskip=.25in]{geometry}

\usepackage{authblk}

\title{Recursive Uncertainty-Gated Image Registration for Learning-based Algorithms}

\author[1]{Clara Rodrigo González}
\author[2]{Oscar Bates}
\author[3]{Fu Siong Ng}
\author[1]{Meng-Xing Tang}

\affil[1]{Department of Bioengineering, Imperial College London, London, UK \\
  \texttt{\{cr418, mengxing.tang\}@imperial.ac.uk}}
\affil[2]{Department of Earth Science and Engineering, Imperial College London, London, UK \\
  \texttt{o.bates1@imperial.ac.uk}}
\affil[3]{National Heart and Lung Institute, Imperial College London, London, UK \\
  \texttt{f.ng@imperial.ac.uk}}

\date{September 2026}

\begin{document}

\maketitle

\begin{abstract}
    Conventional image registration algorithms are robust to domain shifts and achieve low errors, but they are slow and computationally expensive. Deep-learning methods are efficient at inference-time, but face challenges in out-of-domain samples. We propose Recursive Uncertainty-Gated Image Registration (RUGI), an algorithm for iteratively refining deformation fields predicted by learning-based registration models. At each iteration, the registration model predicts an incremental deformation, and a gating map modulates the update. Refinements are hence concentrated in regions that remain difficult to register. We explore two gating strategies: a learned uncertainty-based approach and an image residual error approach. We evaluate RUGI on cardiac MRI and echocardiography datasets and show consistent improvements over single-step inference. Ablation experiments demonstrate that iterative refinement alone improves registration, but informative spatial gating provides a significant additional benefit. The error-gated variant of RUGI can also be applied directly to existing pretrained models; applied to VoxelMorph, TransMorph, and CycleMorph, it yields MSE reductions of 27--37\% with no modification to the original training procedure. The improvements in registration performance are reflected in decreased errors in ejection fraction estimation relative to ground truths. These results demonstrate that spatially selective iterative refinement provides an effective strategy to improve registration accuracy at inference-time.
\end{abstract}

\section{Introduction}

Image registration is a fundamental task in medical image analysis, with applications in diagnostics, treatment planning, and image-guided intervention. In magnetic resonance imaging (MRI), image registration is used to align scans acquired at different time points, contrasts, or respiratory states, enabling longitudinal analysis and multimodal fusion. In ultrasound (US), image registration is used to compensate for tissue motion during freehand acquisitions to improve temporal consistency, and enable downstream analyses such as strain estimation, image compounding, and vascular imaging. 

In recent years, learning-based image registration methods have gained substantial attention due to their speed at inference and their performance. Typically, a neural network receives a fixed and moving image pair and estimates a dense displacement field that can be used to warp one image to match the other. Learning-based approaches enable fast inference at test time and can learn complex, data-driven regularisation priors, and are able to generalize across image pairs and, in some cases, across datasets without a need for further training \cite{balakrishnan_voxelmorph_2019, chen_survey_2025}.

The estimation of spatial correspondence between images is inherently ill-posed: a given pair of images rarely constrains a single unique correspondence field, and the estimated transformation can be highly sensitive to small changes in image content or quality. The degree of ill-posedness is spatially varying across the image pair. Regions with high image quality, clear structure and small, simple motion will tend to be well-constrained, yielding deformations that are unique and stable. Other image regions, with low image quality or complex deformations will constrain correspondences badly, meaning more potential solutions exist. This means the difficulty of the registration task varies across the field of view, and no single fixed strategy is optimal everywhere. This sensitivity to data characteristics limits the ability of fixed, trained models to generalize reliably to new domains, motivating inference-time strategies that can adapt to the specific properties of each image pair.


While conventional registration algorithms perform well and are dataset-agnostic, they are computationally expensive. In contrast, deep-learning methods are quick but are not able to adapt to out-of-domain or challenging samples. We propose an iterative strategy which improves performance of deep-learning models. Our contributions are:
\begin{enumerate}
    \item We introduce a registration refinement framework where iterative updates are gated, concentrating refinement in regions that need it. 
    \item We evaluate two gating strategies, based on predictive uncertainty and residual image error, and show both improve registration accuracy above single-step inference. 
    \item Through an ablation study, we separate the contribution of iterative refinement and spatial gating. 
    \item We show the method can be applied to pretrained models with no need for retraining by using an image similarity-based sampling function. 
    \item We validate RUGI for uncertainty estimation and cardiac functional parameter estimation.
\end{enumerate}

\section{Background}

Iterative and cascaded registration strategies emerged as a way to handle large or complex deformations by decomposing the registration task into a sequence of simpler predictions. LapIRN estimates the deformation in a coarse-to-fine manner using a Laplacian pyramid, where the network at each pyramid level predicts residual deformation fields that are composed with coarser predictions to recover large diffeomorphic deformations \cite{martel_large_2020}. Cascaded VoxelMorph-based approaches similarly improve alignment through a sequence of refinement stages, in which each network receives the partially aligned images from the previous stage and predicts an updated deformation field \cite{xu_4dct_2023}. Related ideas have also appeared in recursive and recurrent registration networks, where repeated prediction-and-warping steps progressively expand the transformations that can be estimated by the model and improve registration accuracy \cite{zhao_recursive_2019,sandkuhler_recurrent_2019,hu_recursive_2022}. These methods apply refinements uniformly across the full image domain, regardless of the spatial heterogeneity or the confidence in the predictions. Moreover, as they introduce changes to model architecture, any further optimisation requires retraining. 

A related line of work for inference-time refinement is test time optimisation (TTO), where the model itself or its outputs are adapted given the test data. \cite{zhu_test-time_2021} constructed a network to model the residual deformations and fine-tune the registration model on each test pair. \cite{aylward_meta-registration_2022} formulated registration as a meta-learning task, training networks to maximise efficiency of test-time optimisation. \cite{ma_iirp-net_2024} proposed a novel architecture which uses iterative multi-scale inference using residual flow estimators. While these methods improve accuracy at inference, they apply refinement within the network architecture, so their application to new out of domain data would require model redesign, specialized training and reoptimisation. As before, refinement is applied across the images without selectivity.

Several works have explored how uncertainty can be incorporated into the registration process itself. Uncertainty-guided approaches have been proposed to weight regularization terms or consistency constraints according to the estimated epistemic uncertainty of the model, thereby allowing uncertain regions to contribute differently during training or optimization \cite{xu_double-uncertainty_2022}. More recently, Tian et al.\ proposed a model-agnostic, inference-time uncertainty estimation framework for pretrained registration networks based on transformation equivariance \cite{tian_uncertainty_2025}. Their method perturbs the input images with known transformations and evaluates the consistency of the predicted deformation under these perturbations. They showed strong correlation between the estimated uncertainty and registration error across multiple anatomical structures and pretrained models.

Although these methods provide important contributions, most require changes to the model structure or training procedure to estimate uncertainty \cite{dalca_unsupervised_2018, gong_uncertainty_2022}. The approaches that use the uncertainty to improve registration do so in limited ways, by changing the regularisation spatiotemporally \cite{xu_double-uncertainty_2022} or by weighting the contribution of different regions to the loss \cite{zhang_heteroscedastic_2024}. \cite{tian_uncertainty_2025} proposed a model-agnostic, inference-time uncertainty estimation framework, but they did not apply the uncertainty estimation to improve the registration itself. To our knowledge, no existing methods leverage the inference-time uncertainty estimation to iteratively refine the deformation field in selective regions. This gap motivates the development of a method that can quantify epistemic uncertainty and use it to guide the focused refinement of registration at inference time, without requiring model retraining. 

\section{Methods}

For a pair of images $I_{mov}(x)$ and $I_{fix}(y)$, the goal is to find a deformation field $\Phi$ which warps $I_{mov}(x)$ to match the underlying structure of $I_{fix}(y)$, such that $x' = x + \Phi$ and $x' \approx y$. In unsupervised motion estimation, a model is trained to predict this deformation field by minimising a loss function between the warped moving image $I_{warped}(x) = I_{mov}(x+\Phi)$ and $I_{fix}(y)$. At inference, this model receives the input images and estimates the deformation field in one step. This deformation field is then used to warp the moving image. 

\subsection{Recursive Uncertainty-Gated Image Registration (RUGI)}
\begin{algorithm}[!h]
\footnotesize
\caption{RUGI Algorithm}\label{alg:uncertainty_gated}
\begin{algorithmic}
\Require Fixed image $I_{\text{fix}}(y)$, moving image $I_{\text{mov}}(x)$, max iterations $T$
\Ensure Composed deformation field $\boldsymbol{\Phi}$
\State $\boldsymbol{\Phi} \leftarrow \text{identity}$
\State $I^{(0)} \leftarrow I_{\text{mov}}(x)$
\State $q_0 \leftarrow \text{Sim}(I_{\text{fix}}(y), I^{(0)})$
\For{$t = 1, 2, \ldots, T$}
    \State $\underline{\phi}^{(t)} \leftarrow R(I_{\text{fix}}(y), I^{(t-1)})$ \Comment{Predict deformation}
    \State $s^{(t)} \leftarrow S(I_{\text{fix}}(y), I^{(t-1)})$ \Comment{Calculate gating} 
    \If{$t = 1$}
        \State $\underline{\hat{\phi}}^{(t)} \leftarrow \underline{\phi}^{(t)}$ \Comment{First step: unmodulated update}
    \Else
        \State $\underline{\hat{\phi}}^{(t)} \leftarrow \underline{\phi}^{(t)} \odot s^{(t)}$ \Comment{Apply gating}
    \EndIf
    \State $\boldsymbol{\Phi}_{\text{new}} \leftarrow \boldsymbol{\Phi} \circ \underline{\hat{\phi}}^{(t)}$ \Comment{Tentative composition}
    \State $I^{(t)} \leftarrow I_{\text{mov}}(x+\boldsymbol{\Phi}_{\text{new}}) $ \Comment{Warp with new field} 
    \State $q_t \leftarrow \text{Sim}(I_{\text{fix}}(y), I^{(t)})$
    \If{$q_t > q_{t-1} $} \Comment{Insufficient improvement, stop}
        \State \textbf{break}
    \EndIf
    \State $\boldsymbol{\Phi} \leftarrow \boldsymbol{\Phi}_{\text{new}}$ \Comment{Accept update}
\EndFor
\State \textbf{Return } $\boldsymbol{\Phi}$
\end{algorithmic}
\end{algorithm}

\begin{figure}[H]
    \centering
    \includegraphics[width=0.85\linewidth]{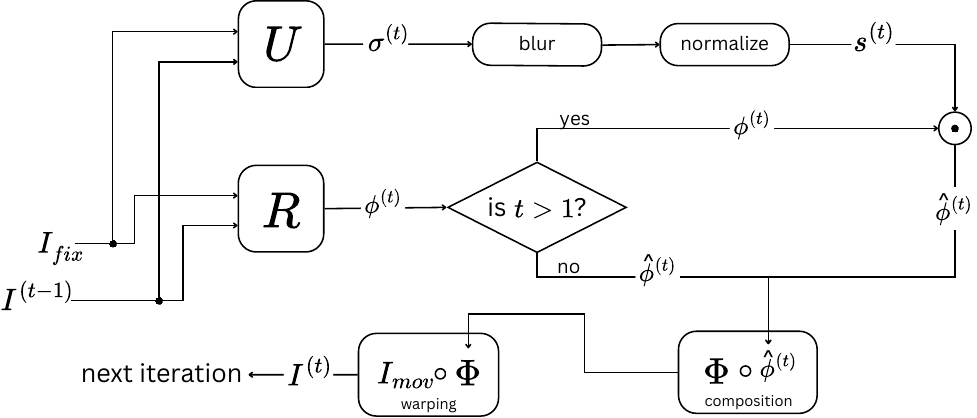}
    \caption[RUGI Flowchart]{\textbf{RUGI Flowchart.} At an arbitrary time step $t$, both the registration network $R$ and the uncertainty network $U$ receive the fixed image and latest warped moving image $I^{(t-1)}$.  The uncertainty network produces a gating map $\sigma^{(t)}$ which, after blurring and normalization, is used to weight registration updates.}
    \label{fig:algo_diagram}
\end{figure}

Algorithm \ref{alg:uncertainty_gated} describes the Recursive Uncertainty-Gated Image Registration (RUGI) Algorithm. At each iteration $t$, the registration model $R()$ predicts a dense deformation field over the full image domain, conditioned on the fixed image $I_{fix}(y)$ and the moving image warped at the previous step, $I^{(t-1)}$. The model outputs a vector  field for that time point $\underline{\phi}^{(t)}=R(I_{fix}(y), I^{(t-1)}) $. In parallel, the gating function $S()$ produces a spatially varying gating map $s^{(t)} = S(I_{\text{fix}(y)}, I^{(t-1)} )$, which is normalized for each image. The gating map modulates the magnitude of the incremental deformation update at each iteration. Figure \ref{fig:algo_diagram} shows a flowchart of Algorithm \ref{alg:uncertainty_gated}.

We used a maximum number of iterations $T=30$. From the second step, the predicted gating was spatially smoothed using a fixed $3\times3$ Gaussian convolution ($\sigma=2$ pixels). The smoothed map was then independently min-max normalised for each sample before being multiplied element-wise with the predicted incremental deformation field. 

Iterative refinement is applied until a similarity-based stopping criterion is satisfied. The similarity metric can be tuned to the specific dataset, or additional stopping criteria can be applied to constrain the deformations. We investigated multiple similarity metrics and empirically chose the MSE, as it was robust and consistently led to reliable results. We added a second stopping criterion as a safeguard against severe folding. After a two-iteration warmup, the Jacobian determinant of the composed candidate field was calculated within the inner region of the field of view. If the 1st percentile of the Jacobian distribution was below -0.4, the candidate update was rejected and iterative refinement was terminated. The calculation of the Jacobian excluded a 5-pixel border around the images. This criterion was used to prevent severe folding rather than to impose strict diffeomorphism. 

The selective gating is anisotropic, and is calculated and applied independently along each axis $s_x^{(t)}$ and $s_z^{(t)}$. The sampled deformation field is then given by:

\begin{equation}
    \underline{\hat{\phi}}^{(t)} = \begin{pmatrix} s_x^{(t)} \odot \underline{\phi}_x^{(t)} \\ s_z^{(t)} \odot \underline{\phi}_z^{(t)} \end{pmatrix}
    \label{eq:anisotropic_sampling}
\end{equation}

\noindent where $\underline{\phi}_x^{(t)}$ and $\underline{\phi}_z^{(t)}$ are the predicted flow components along each axis, and $s_x^{(t)}, s_z^{(t)}$ are the corresponding normalised gating maps. This anisotropic formulation allows the method to modulate registration uncertainty independently per direction, which is particularly important in settings where noise and artefact characteristics differ across axes.


\subsubsection{Uncertainty-based gating}
\label{sec:unc_sampling}
Uncertainty can be used to guide the selective gating process. In this case, regions with low predicted uncertainty receive smaller updates and stabilize earlier in the iterative refinement, while high-uncertainty regions continue to receive updates. 

To estimate uncertainty, we train a simple U-Net $U$ to predict the uncertainty in the deformation domain from the warped and fixed images. During training, the registration model $R$ is sampled $N$ times for each image pair, producing a distribution of deformation fields $\phi_{1..N}$ and warped images $I_{warped}$. The similarity loss between $I_{fix}(x)$ and $I_{warped}$ is used to train the registration model $R$. The per-pixel standard deviation of these fields $\sigma$ is used to train the uncertainty model $U(I_{warped},I_{fixed})$ in a supervised manner using the mean squared error (MSE) as the loss function.

At inference, the uncertainty model predicts
\begin{equation}
\sigma^{(t)}=U(I^{(t-1)}, I_{fix}(x))
\end{equation}
This uncertainty map is then smoothed and normalised to obtain the gating map $s^{(t)}$. 

The uncertainty-based method is predictive and spatially adaptive, and while it requires an additional network, the costs are minimized by the small size of the network (465,138 parameters).

\subsubsection{Error-based gating}
Uncertainty estimates are often evaluated using their correspondence with prediction error \cite{tian_uncertainty_2025, lakshminarayanan_simple_2017}, with high-uncertainty regions being expected to exhibit larger registration errors. Motivated by this relationship, we also consider an error-based weighting strategy, which provides a simpler and cheaper alternative to the uncertainty-based method. 

In this case, a per-pixel error measure between the fixed image $I_{fix}(x)$ and the current warped image $I^{(t-1)}$ is calculated, smoothed and normalized to produce the gating map $s^{(t)}$. Regions with larger mismatch are assigned larger updates and continue to be refined, while regions with smaller residual error are assigned smaller updates.

This error-based gating method uses only the registration model $R$ and does not require retraining when changing datasets. It is computationally inexpensive and easily adaptable, as the residual measure can be replaced to fit new modalities or settings. 

A post hoc measure of the uncertainty can be calculated using the standard deviation of the incremental deformation estimates across iterations. However, this quantity should not be interpreted as a calibrated uncertainty estimate; rather, it provides a measure of spatial heterogeneity in the refinement process.

\subsection{Training Procedure}

\begin{figure}[H]
    \centering
    \includegraphics[width=0.8\linewidth]{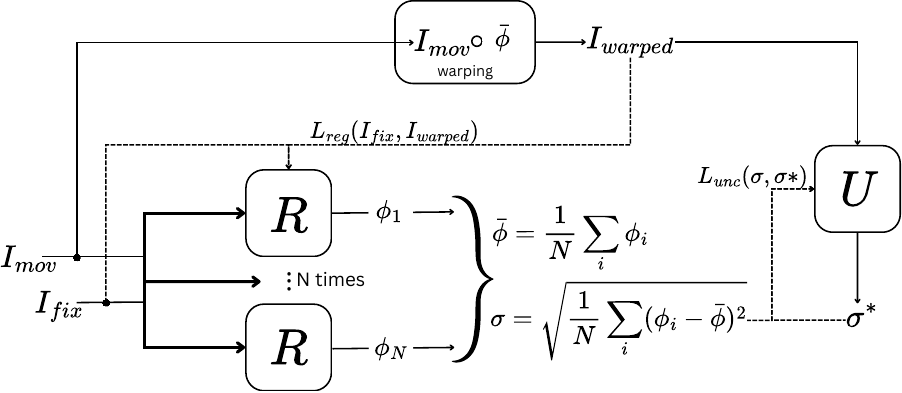}
    \caption[Training Flowchart for uncertainty-based RUGI]{\textbf{Training Flowchart for uncertainty-based RUGI}. The registration network $R$ is sampled N times to estimate the mean $\bar{\phi}$ and standard deviation $\sigma$ of the estimated deformation field. The standard deviation $\sigma$ is then used to train the uncertainty network $U$ in a supervised manner. }
    \label{5fig:training}
\end{figure}

While RUGI works at inference time with no modification to the registration model, the uncertainty estimation requires a lightweight network to be trained. Both networks are trained concurrently, so the uncertainty network learns to predict uncertainty specific to $R()$'s predictions.  Figure \ref{5fig:training} shows a flowchart of the training procedure for the uncertainty-based RUGI. During training, the registration model $R()$ is sampled $N$ times for each image pair, producing a distribution of deformation fields $\phi_{1..N}$. The registration model was sampled $N=5$ times for each training batch. The per-pixel standard deviation of these fields $\sigma$ is used to train the uncertainty model $U$ in a supervised manner using the mean squared error (MSE) as the loss function. A stochastic deformation sample from $R$ is used to warp $I_{mov}(x)$, yielding $I_{warped}(x')$. The similarity loss between $I_{fix}(y)$ and $I_{warped}(x')$ is used to train the registration model $R$. The full loss function during training used the MSE as the similarity term, and an added bending energy term as the regularisation loss. At inference, the uncertainty model predicts $\sigma^{(t)}=U(I^{(t-1)},I_{fix}(y))$. This uncertainty map is then smoothed and normalized to obtain the gating map $s^{(t)}$. While this requires an additional network, the costs are minimized by the small size of the network, which only has 465k parameters. 

The registration and uncertainty networks were trained until convergence using a batch size of 20 and separate AdamW optimisers with a weight decay of \(4\times10^{-5}\). For the ACDC and CAMUS experiments, both networks used a learning rate of \(10^{-3}\). The registration objective consisted of the image MSE, scaled by a factor of 100, together with bending-energy regularisation weighted by \(4\times10^{-4}\). The uncertainty network was trained using MSE between its predicted map and the min--max-normalised pixelwise standard deviation obtained from five stochastic samples of the Bayesian registration network. A fixed random seed was used for reproducibility.

Since $U()$ estimates epistemic uncertainty, high values indicate regions where the model's predictions are inconsistent across multiple passes. A BNN trained on image pairs will exhibit high predictive variance in regions where the deformation is difficult to estimate consistently, due to its magnitude, complexity, or image quality. Conversely, regions of background or static anatomy are registered consistently, yielding low $\sigma$.

\subsection{Implementation details}
\paragraph{Registration model} We used a compact Bayesian U-Net-inspired architecture containing two downsampling stages and eight convolutional layers, with approximately 1.9M trainable parameters. All input images were resized to 128x128 pixels. This architecture was selected because the predicted deformation fields were expected to vary smoothly over spatial scales, reducing the need for deep networks. This architecture was found to obtain accurate registration across the evaluated ultrasound and MRI datasets. The BNN provides a distribution over predicted deformation fields, which is useful for estimating spatial uncertainty. At inference, the stochasticity is turned off. The Bayesian layers were parameterised using prior standard deviations \(\sigma_1=1.0\) and \(\sigma_2=0.002\), mixture coefficient \(\pi=0.5\), posterior mean \(\mu=0\), and posterior parameter \(\rho=-3.0\).

\paragraph{Uncertainty model} The uncertainty model should be quick, ideally with a cost comparable or lower to that or the registration itself, and only use the image pair rather than the deformation field. Since our method is iterative, we require an uncertainty estimate that can be recomputed at each refinement step. For these reasons, we chose to use a lightweight U-Net network to approximate the empirical uncertainty. The network consists of 2 layers and 465,138 parameters.

\paragraph{Error-based sampling} The residual error can be calculated using a range of similarity measures. We chose the local normalised cross-correlation for its robustness across imaging modalities and its insensitivity to global intensity differences. To increase the dynamic range of the sampling map and amplify the contrast between high- and low-error regions, an exponential transformation is applied to the residual map prior to normalisation.


\paragraph{Datasets}
We evaluated our selective gating algorithm on two datasets. First, the Automatic Cardiac Diagnosis Challenge (ACDC) dataset, which contains cardiac magnetic resonance acquisitions from 1.5T and 3.0T MRIs \cite{bernard_deep_2018}. The total training dataset has 100 patients with two volumes per patient, and 560 training image pairs. We extracted axial cardiac images to perform this analysis in two-dimensional images. Cardiac Acquisitions for Multi-structure Ultrasound Segmentation (CAMUS) is an ultrasound dataset containing cardiac acquisitions from 450 healthy and anomalous patients in multiple views \cite{leclerc_deep_2019}. Each acquisition was segmented by a clinical expert, providing ground truth anatomical information. 4500 images were used for training. We used each patient as an independent sample. Train:test:validation splits were performed by dividing across patients so no patient was in both.

\subsection{Ablation Study}
An ablation study was performed to determine which components of the proposed algorithm are responsible for the improvement in registration accuracy. The full method using both uncertainty-based and error-based gating was compared to three ablations. 
\begin{itemize}
    
    \item A1 no gating. The iterative multistep registration is applied without any spatial gating or modulation. The full predicted update is applied at each iteration. This ablation tests for improvements due to the iterative registration itself.
    \item A2 random gating. The gating map is replaced with a random gating map, which is then smoothed and normalized. This ablation tests for the improvements due to spatially varying gating. 
    \item A3 intensity-based gating. The moving image is smoothed with a gaussian blur and normalized to yield the gating map. This ablation tests whether a simpler image-based structural prior is sufficient to guide refinement. 
\end{itemize}

\subsection{Application to pretrained models} 
An important advantage of the proposed error-based gating strategy is that it can be applied entirely at inference time, without modifying the original training procedure. This allows the method to be used as a plug-in iterative refinement framework for existing pretrained registration models.

To validate this, we applied the method to several publicly available pretrained models, namely TransMorph, VoxelMorph, CycleMorph, with pretrained weights extracted from \cite{tian_unigradicon_2024}. These models were trained on the IXI dataset, a 3-D brain MRI registration dataset \cite{noauthor_ixi_nodate}. For each model, we compared the original one-step inference result with the output obtained when the same model was used within our iterative error-based gating framework.
 
This experiment was designed to assess whether the proposed method can improve registration accuracy without retraining, and whether these gains generalize across different network architectures. All pretrained-model experiments were evaluated using the same metrics as in the main experiments.

\subsection{Evaluation}
Registration performance was evaluated on two held-out test cohorts, of 50 patients on ACDC and 100 patients on CAMUS, with splits dividing by patient so no patient is in both the training and test cohorts. Unless otherwise stated, the experiments were performed using a fixed random seed. 

Registration accuracy was evaluated between the fixed $I_{fix}(y)$ and warped $I_{warped}(x')$ images using the mean squared error (MSE), normalised cross-correlation (NCC), mutual information (MI) and structural similarity index measure (SSIM). All statistical analyses were carried out at the patient level. Where multiple measurements were available for a patient, they were aggregated before statistical testing. Pairwise comparisons between baselines and RUGI were performed using two-sided Wilcoxon signed-rank tests. Effect sizes were quantified using paired rank-biserial correlation, with differences oriented such that positive values favoured RUGI. Benjamini--Hochberg (BH) false-discovery-rate correction was applied within each family of multiple corrections, with w$p_{BH}<0.05$ considered statistically significant. 

For the ablation study, the similarity metrics were combined using the mean signed-rank effect $G$, with significance assessed using a patient-level sign-flip permutation test with 100,000 permutations and BH correction. 

\subsection{Validation of Estimated Uncertainty}

To assess whether the uncertainty predicted by our additional network identifies regions with residual registration error, registration was compared spatially with the absolute error $|I_{fixed}-I_{warped}|$. In the ultrasound datasets, this error map was lightly smoothed to decrease the influence of speckle intensity variations. Pixels with negligible intensity in both fixed and warped images were excluded.

For each image, valid pixels were ranked independently by predicted uncertainty and residual error. The top 20\% of pixels for each measure were identified. Spatial overlap was calculated as the proportion of the top-20\% uncertainty pixels that were also present in the top-20\% residual error. As both masks contain 20\% of the valid pixels, the expected overlap under random spatial correspondence is 20\%. Spatial enrichment was calculated as the observed overlap divided by 0.20, such that a value of 1 corresponds to the expected overlap due to chance. 

To account for small spatial offsets between uncertainty and residual-error regions, we calculated tolerance-aware overlap. For a tolerance of $r$ pixels, each binary top-20\% mask was dilated using a square neighbourhood extending $r$ pixels in each direction. The proportion of uncertainty-mask pixels falling within this neighbourhood was calculated, together with the reciprocal proportion of residual-error pixels falling within that region. The mean of these two values was reported as the tolerance-aware overlap. Tolerance values $r$ of 1, 2, 3 and 5 pixels were evaluated.

\subsection{Functional Parameters}
\label{sec:cimethod}

To assess whether the deformation fields predicted by RUGI capture the full range of cardiac motion, we derived a registration-based surrogate of LVEF between end-diastole (ED) and end-systole (ES). For each ED-ES pair, the registration model was used to estimate the motion from ES to ED. The estimated transformation is used to warp the ES left-ventricular mask into the ED domain. EF was then computed from the relative change in chamber size: 
$$EF_{reg} = 100 \times (A_{ES,warped}-A_{ES})/A_{ES,warped}, $$ 
where 
$$A_{ES,warped}=Area(warp(M_{ES}, \mathbf{\Phi}_{ES\rightarrow ED})),$$

and $A_{ES}$ and $A_{ED}$ are the areas of the left-ventricular masks in end systole and diastole, respectively. $\mathbf{\Phi}_{ES\rightarrow ED}$ is the transformation from ES to ED. With perfect registration, $A_{ES,warped}$ approaches $A_{ED}$, and this EF calculation should match the EF estimated using ground truth masks. In datasets with 2D masks, this yields an area-based surrogate of EF rather than a true volumetric clinical EF, but still provides a functional measure of how well the estimated deformation captures cardiac contraction.

For the RUGI variants, per-sample uncertainty intervals (UI) are derived for all parameters by Monte Carlo perturbation of the deformation field, using the per-pixel epistemic uncertainty map to scale the perturbations. Let $\phi$ denote the point-estimate deformation field, and $\hat{\sigma(\mathbf{x})}$ the per-pixel uncertainty map. For each patient, 50 perturbed fields were sampled through $$\phi^{(s)} = \phi + \epsilon^{(s)}, \epsilon \sim \mathcal{N}(0,\hat{\sigma}(\mathbf{x})^2)$$
where the noise is first drawn independently at each pixel and then smoothed using a Gaussian kernel. This spatial smoothing reintroduces local correlation between pixels, ensuring the produced fields are spatially correlated. Each perturbed field $\phi^{(s)}$ is used to calculate the clinical parameters, providing a measure of uncertainty. The 95\% intervals in Table \ref{tab:cardiac_results} are the 2.5th and 97.5th percentiles of this distribution. 


\section{Results}
The results are structured as follows. Section \ref{results_reg} assesses the function of RUGI as a registration framework. This includes its improvements in performance on new and pretrained models, the ablation study, and a computational cost analysis. Section \ref{results_unc} validates the uncertainty estimates generated by RUGI-Unc, and their potential utility for estimation of downstream functional parameters. 

\subsection{RUGI For Registration}
\label{results_reg}
\begin{figure}
    \centering
    \includegraphics[width=\textwidth]{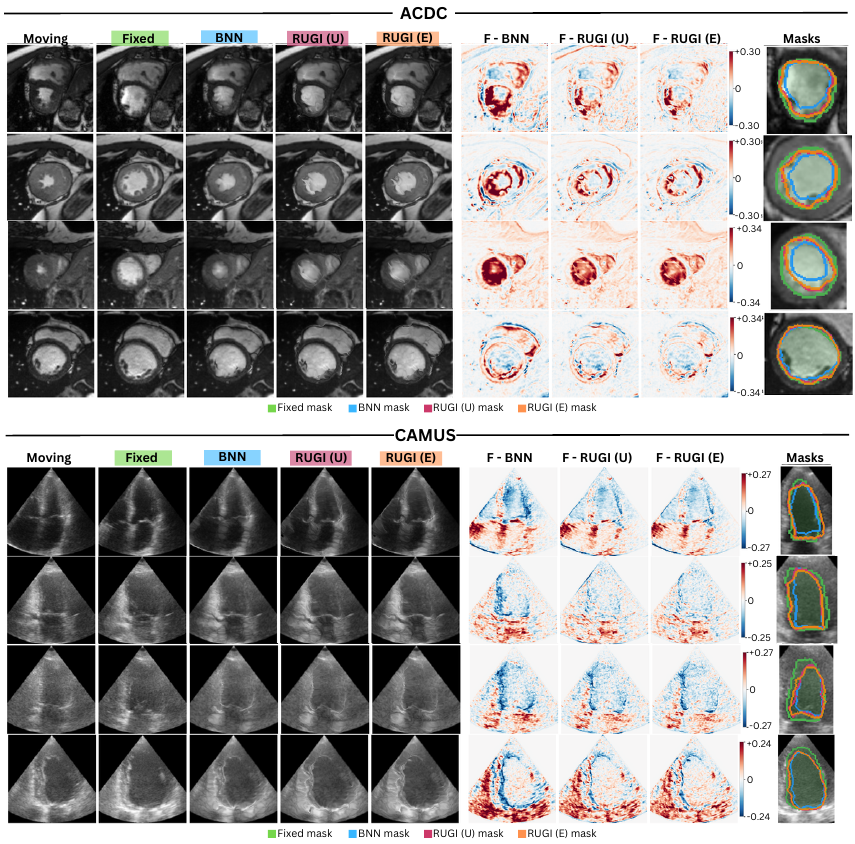}
    \caption[Example results using baseline and RUGI]{\textbf{Example results using baseline and RUGI.} Results were obtained using the proposed method with uncertainty-based (U) and error-based (E) gating, in comparison to the baseline BNN. Example warped images, error maps and warped masks are shown for the two datasets. }
    \label{fig:examples}
\end{figure}

\begin{figure}[H]

  \includegraphics[width=\linewidth]{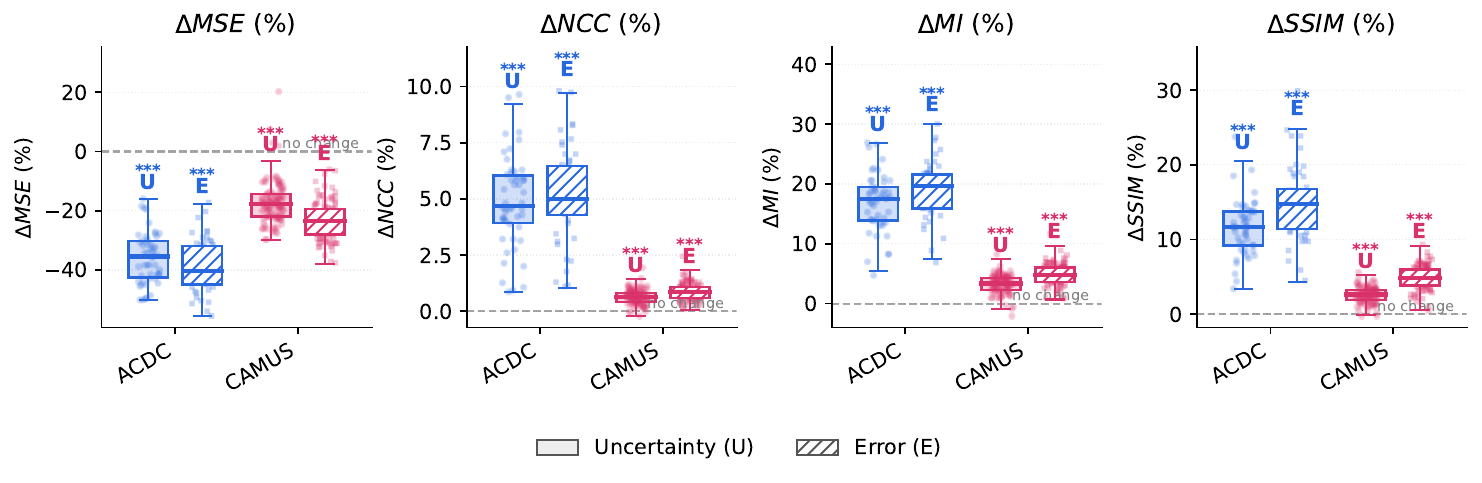}
  \caption[Image similarity improvements using RUGI]{\textbf{Image similarity improvements using RUGI.} Percentage change in image similarity metrics relative to one-step method across the two datasets. Our proposed methods are shown using both uncertainty-based (U, solid bars) and error-based (E, dashed bars) gating. }
  \label{fig:ratio_results}
\end{figure}

Figure \ref{fig:examples} shows example registrations across the two datasets for our proposed methods, as well as a baseline single-step BNN. Figure \ref{fig:ratio_results} shows the improvement according to the four metrics as a ratio of the performance of the one-step inference. Both proposed RUGI variants improved image similarity relative to the single-step baseline across both datasets. Patient-level statistical analyses compared paired measurements between the single-step and recursive methods. This was done using two-sided paired Wilcoxon signed-rank tests with Benjamini-Hochberg false-discovery-rate correction. Both RUGI-Unc and RUGI-Err produced significant improvements in all four metrics. These improvements were consistent across patients, with RBC effect sizes close to 1 for most comparisons. Significance is maintained even after correction. 

To further validate the improvements in registration quality beyond intensity-based metrics, we evaluated anatomical alignment directly using the Dice similarity coefficient (DSC) for each patient. Deformation fields produced by each method were used to warp the segmentation masks, and DSC was computed between the warped and fixed masks. The DSC was averaged within each patient, so one patient constituted one independent observation. Both proposed methods achieved a statistically significant improvement in DSC relative to the single-step  baseline (paired two-sided Wilcoxon signed-rank test) after Benjamini--Hochberg correction. In ACDC, the uncertainty-based method increased the median DSC across all tissues from $73.8\%$ to $83.7\%$, and the error-based method increased it to $84.6\%$. The median paired improvements were $+9.8\% [+7.2, +13.1]$ with RUGI-Unc ($RBC=1.00$, $p_{BH}<0.0001$) and $+9.7\% [+7.5, +14.7]$ with RUGI-Err ($RBC=1.00$, $p_{BH}<0.0001$). On CAMUS, RUGI-Unc increased the median DSC from $78.2\%$ to $82.3\%$ , while RUGI-Err achieved a median of $82.0$. Median improvements were $+3.9\% [+3.5, +4.4]$ for RUGI-Unc ($RBC=1.00$, $p_{BH}<0.0001$) and $+3.7\% [+3.2, +4.3]$ for RUGI-Err ($RBC=0.981$, $p_{BH}<0.0001$). These results demonstrate that the improvements observed in intensity-based image similarity metrics correspond to a corresponding but variable gain in anatomical alignment.

\subsubsection{Ablation Study}

\begin{figure}[]
    \centering
    \includegraphics[width=0.95\linewidth]{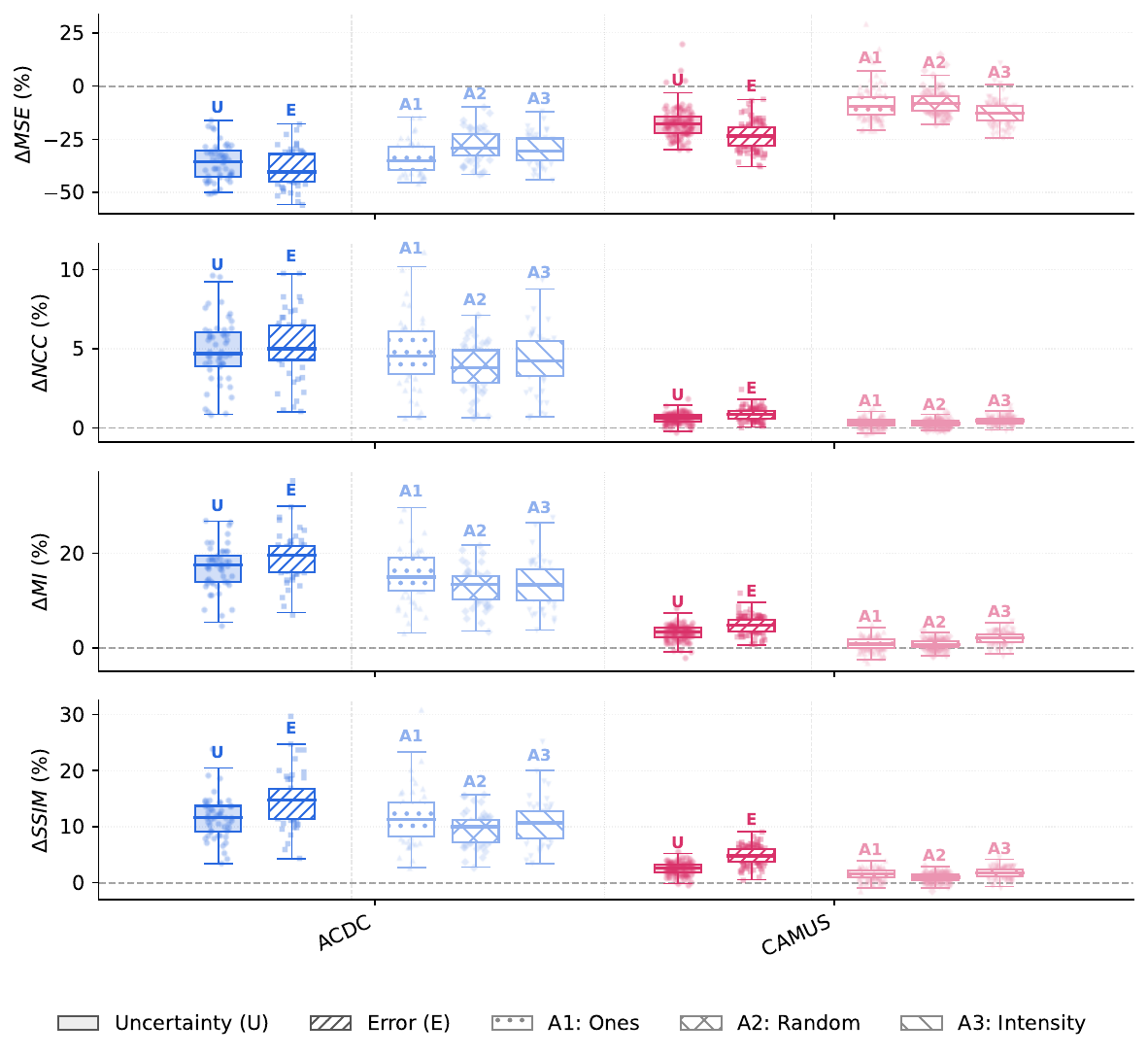}
    \caption[Ablation study with respect to image similarity]{\textbf{Ablation Study}. Improvement ratios relative to one-step BNN baseline in the two datasets using our proposed method as well as the three proposed ablations. A1 shows full steps with no gating, A2 shows gating using a random blurred gating, A3 used gating based on the blurred intensity of fixed and moving images. }
    \label{fig:ablation}
\end{figure}

\begin{table}[]
\centering
\footnotesize
\begin{tabular}{llcc}
\toprule
\textbf{Method} & \textbf{Ablation} &
\textbf{ACDC} & \textbf{CAMUS} \\
\midrule

\multirow{3}{*}{RUGI-Unc}
& A1: Ones      & $+0.40^{**}$  & $+1.00^{***}$ \\
& A2: Random    & $+0.99^{***}$ & $+1.00^{***}$ \\
& A3: Intensity & $+0.88^{***}$ & $+0.98^{***}$ \\

\cmidrule{1-4}

\multirow{3}{*}{RUGI-Err}
& A1: Ones      & $+0.93^{***}$ & $+1.00^{***}$ \\
& A2: Random    & $+1.00^{***}$ & $+1.00^{***}$ \\
& A3: Intensity & $+1.00^{***}$ & $+1.00^{***}$ \\

\bottomrule
\end{tabular}

\caption[Global statistical analysis of ablation study]{
\textbf{Global statistical analysis of the ablation study.}
Cells show the global signed-rank effect $G$, combining MSE, NCC, MI and SSIM with equal rank-based weighting. Positive values indicate better overall performance of the RUGI method relative to the corresponding ablation, with $G=1$ indicating that all ranked paired differences favour RUGI. Significance was assessed using a patient-level synchronised sign-flip permutation test and Benjamini--Hochberg correction across the six RUGI-versus-ablation comparisons within each dataset: $^{*}p_{\mathrm{BH}}<0.05$, $^{**}p_{\mathrm{BH}}<0.01$, $^{***}p_{\mathrm{BH}}<0.001$; \textit{ns} = not significant. Each patient was taken as an independent measurement (ACDC: $n=50$, CAMUS: $n=100$.) }
\label{tab:ablation}
\end{table}

Figure \ref{fig:ablation} compares the two proposed RUGI variants with three ablated gating strategies. The ablation study was designed to determine whether the spatial gating signal contributes additional value beyond iterative refinement alone. RUGI-Unc and RUGI-Err were therefore compared with a constant no-gating schedule (A1), a random gating schedule (A2), and an intensity-based gating schedule (A3), across two datasets and four image-similarity metrics.

To test the overall effect of the gating strategy, MSE, NCC, MI and SSIM were analysed jointly using a patient-level synchronised sign-flip permutation test. For each metric, paired differences were oriented such that positive values favoured the proposed RUGI method and converted to signed-rank effects; the global statistic $G$ was then calculated as the mean across the four metrics. Thus, $G>0$ indicates an overall advantage of RUGI over the corresponding ablation, while $G=1$ indicates complete directional agreement across the ranked paired differences. Benjamini-Hochberg false-discovery-rate correction was applied across the six RUGI-versus-ablation comparisons within each dataset. Table \ref{tab:ablation} shows the statistical results from this study across the two proposed RUGI variants and two datasets. 

Both proposed RUGI variants significantly outperformed the ablations in almost all comparisons. For RUGI-Unc, the largest effect sizes were found against the random gating, with $G=0.99, 1.00$ on ACDC and CAMUS, respectively. RUGI-Unc also significantly outperformed the intensity-based gating in ACDC and CAMUS, $G=0.88, 0.98$, respectively. Compared to the ungated iterative strategy (A1), RUGI-Unc remained significantly better on ACDC ($G=0.40, p_{BH}=0.003$) and CAMUS ($G=1.00, p_{BH}<0.001$). 

RUGI-Err showed an even more consistent separation from the ablations. It significantly outperformed A1, A2 and A3 on every dataset after multiple-comparison correction. Global effects were large, ranging from \(G=0.93\) to \(1.00\) on ACDC, and \(G=1.00\) for all three comparisons on CAMUS. For almost all RUGI-versus-ablation comparisons, all four constituent image-similarity metrics favoured the proposed method, demonstrating that the global effects were not driven by a single metric. The only exception was RUGI-Unc versus A1 on ACDC, for which three of the four metrics favoured RUGI-Unc overall.

Importantly, the ablated gating strategies improved image similarity relative to single-step registration. This indicates the iterative refinement of deformation fields is beneficial even without selective gating. However, this repeated iteration alone does not fully explain the performance of RUGI. The inferior performance of the random and intensity-based gating also indicates the gating must contain relevant information to the residual registration error. 

Taken together, the ablation results demonstrate two mechanisms by which our proposed framework improves performance. Recursive optimisation of the deformation fields provides an improvement over single-step inference, while selective gating provides a significant additional improvement. Both uncertainty and error estimation provide informative spatial gating, while uninformed schedules consistently and significantly underperform regardless of dataset or metric.

\subsubsection{Application to pre-trained models}
\begin{figure}
    \centering
    \includegraphics[width=0.95\linewidth]{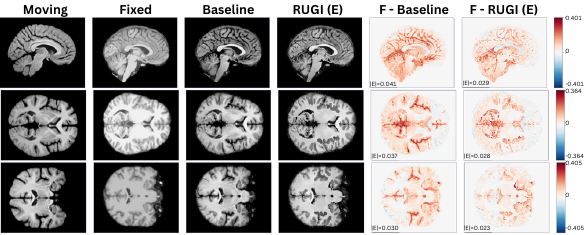}
    \caption{Example warped images and residual errors obtained using pretrained TransMorph with and without RUGI.}
    \label{fig:pretrained_examples}
\end{figure}

\begin{figure}[!h]
    \centering
    \includegraphics[width=\linewidth]{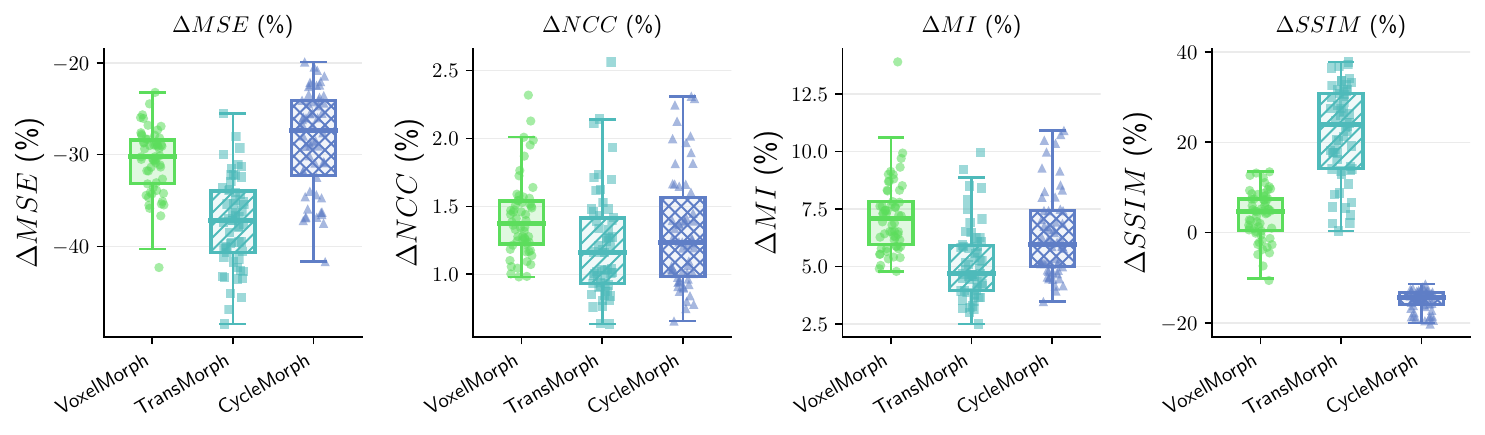}
    \caption[RUGI on pretrained models]{\textbf{RUGI on pretrained models.} Similarity metrics as a percentage change from the single-step approach. All three pretrained models obtained significant improvements in MSE after our focused gating inference was applied. VoxelMorph got a reduction of 30.2\% in MSE, TransMorph got 37.1\% and CycleMorph 27.4\%. }
    \label{fig:pretrained_models}
\end{figure}

The focused gating algorithm was applied at inference to three pretrained registration models, VoxelMorph, TransMorph, and CycleMorph, evaluated on 58 paired samples. Examples of the registration performance with and without RUGI using TransMorph are shown in Figure \ref{fig:pretrained_examples}, and the effect of RUGI in image similarity relative to the baseline single-step inference is shown in Figure \ref{fig:pretrained_models}. RUGI significantly improved MSE, MI and NCC across all three architectures according to a two-sided Wilcoxon signed-rank test after Benjamini--Hochberg ($p_{BH}=3.82 \times10^{-11}$ for each comparison). 

Median MSE was reduced by 30.2\%, 37.1\%, and 27.4\% in VoxelMorph, TransMorph, and CycleMorph, respectively, with $RBC=1.00$ for each case, indicating every non-zero ranked paired difference benefited RUGI. NCC increased by $1.37\%$, $1.16\%$ and $1.23\%$, respectively, again with $RBC=1.00$. MI also increased consistently, by $7.07\%$, $4.68\%$ and $5.96\%$ for VoxelMorph, TransMorph and CycleMorph, respectively ($RBC=1.00$).

SSIM changed less consistently. VoxelMorph improved by a median $4.68\%$ ($RBC=0.646$,$p=1.89\times10^{-5}$), TransMorph improved by $23.9\%$ ($RBC=1.00$, $p_{BH}=3.82\times10^{-11}$). In contrast, CycleMorph obtained a significant decrease in SSIM of $14.4\%$ ($RBC=-1.00$, $p_{BH}=3.82\times10^{-11}$). 

Overall, RUGI-Err produced highly consistent improvements in MSE, NCC and MI across all three pretrained networks without retraining, while its effect on SSIM was architecture-dependent.

\subsubsection{Computational Cost}

\begin{table}[H]
\centering
\footnotesize
\begin{tabular}{lcccccc}
\toprule
\textbf{Dataset} & \textbf{BNN (s)} & \textbf{RUGI-Unc (s)} & \textbf{RUGI-Err (s)} & \textbf{Steps (U)} & \textbf{Steps (E)} & \textbf{Ratio (U/O)} \\
\midrule
ACDC & 0.014 & 0.141$^{***}$ & 0.067$^{***}$ & 8.0 & 4.0 & 10.0$\times$ \\
CAMUS & 0.014 & 0.073$^{***}$ & 0.104$^{***}$ & 4.0 & 7.0 & 5.2$\times$ \\
\bottomrule

\end{tabular}
\caption[Computational cost for RUGI]{\textbf{Computational cost for RUGI.} Time is median wall-clock inference time per sample (seconds). Steps is the median number of registration steps taken. Ratio = multistep / one-step time. Significance (BH-corrected Wilcoxon): $^{*}p{<}0.05$, $^{**}p{<}0.01$, $^{***}p{<}0.001$.}
\label{tab:compute}
\end{table}

Table \ref{tab:compute} reports the median inference time per sample and median number of registration steps for both proposed methods relative to the single-step BNN baseline. The multistep inference is significantly slower than the one-step baseline in all cases (BH-corrected Wilcoxon, $p < 0.001$ throughout), which is expected given that it performs multiple forward passes. The uncertainty method takes between $5.2\times$ and $10.0\times$ longer than the baseline depending on the dataset. The error-based method is faster than the uncertainty-based method in ACDC, taking less steps before convergence. In CAMUS, the error-based method is slower, potentially due to the higher noise in the ultrasound dataset making the error proxy less robust. The variation in stopping criterion across datasets and methods shows the adaptability of our method to different datasets and samples. 

\begin{figure}[H]
     \centering
     \begin{subfigure}[b]{0.58\textwidth}
         \centering
         \includegraphics[width=\textwidth]{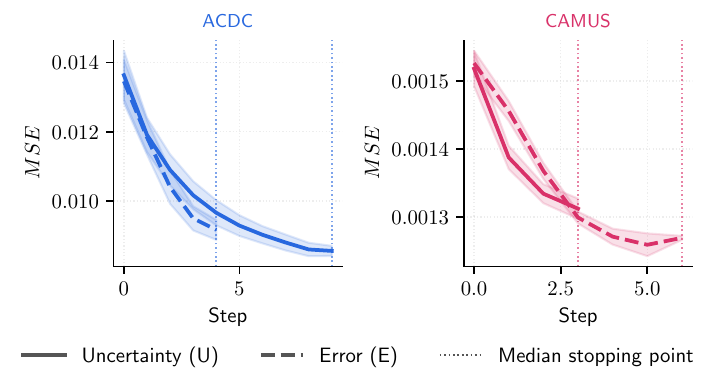}
         \caption{}
         \label{fig:cost_scatterplot}
     \end{subfigure}
     \hfill
     \begin{subfigure}[b]{0.4\textwidth}
         \centering
         \includegraphics[width=\textwidth]{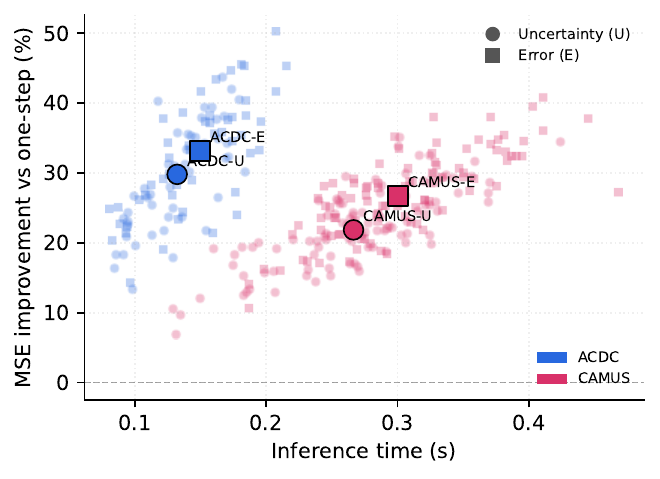}
         \caption{}
         \label{fig:five over x}
     \end{subfigure}
        \caption[Computational cost analysis]{\textbf{Computational cost analysis}. Both methods improve monotonically with each step and achieve consistent MSE reductions relative to the one-step baseline. \textbf{a)} Mean loss per iteration across 100 samples for uncertainty-based (solid line) and error-based (dashed line) methods (shading = $\pm 1$ std). Dotted vertical lines indicate the median stopping point per method. Loss decreases monotonically in all cases. \textbf{b)} Quality--compute trade-off. Five independently initialised models were trained for each dataset. Points show per-patient values, while larger points show the average per-method for each dataset. Lower and further left shows improved performance and higher computational efficiency.}
        \label{fig:compute}
\end{figure}

Figure~\ref{fig:compute}(a) shows the mean loss trajectory across 100 samples per dataset, with shading indicating one standard deviation, and Figure~\ref{fig:compute}(b) shows the quality-compute trade-off . On ACDC, RUGI-Err converges earlier and leads to slightly larger residual error, while RUGI-Unc requires more steps but leads to lower median errors. On CAMUS, RUGI-Err converges after more steps than RUGI-Unc and leads to lower residual errors. For both proposed methods, the loss decreases monotonically from step~0 to the median stopping point (dotted vertical line), confirming that the iterative gating procedure makes consistent improvements at each step rather than oscillating. The quality-compute trade-off shows that, although RUGI results in longer inference times relative to the single-step baseline, this increase in compute has corresponding improvements in performance in both datasets. 

\subsection{RUGI For Uncertainty Estimation}
\label{results_unc}

\subsubsection{Validation of Estimated Uncertainty}

\begin{figure}[H]
    \centering
    \includegraphics[width=0.95\textwidth]{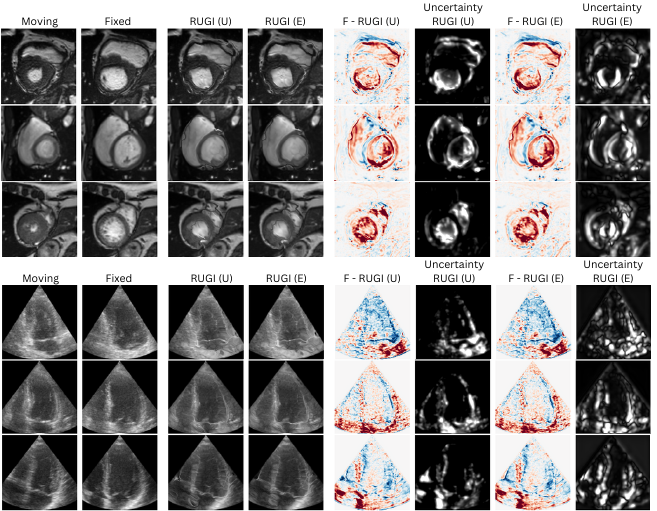}
    \caption[Example uncertainty maps.]{\textbf{Example uncertainty maps.} Warped images are shown for RUGI-Unc (U) and RUGI-Err (E), alongside the corresponding fixed–warped residuals. For RUGI-Unc, the uncertainty map represents the predicted epistemic uncertainty of the registration model. For RUGI-Err, the displayed map represents the post-hoc variability of incremental deformation estimates across refinement iterations. }
    \label{fig:sampling_examples}
\end{figure}

Figure \ref{fig:sampling_examples} shows example outputs for our proposed methods, as well as their uncertainty maps. For RUGI-Unc, the uncertainty map represents the predicted epistemic uncertainty of the registration model. For RUGI-Err, the displayed map represents the post-hoc variability of incremental deformation estimates across refinement iterations. RUGI-Unc generally produces smoother maps, with elevated values concentrated around myocardial structures and tissue boundaries, whereas RUGI-Err produces finer-scale spatial structure that is less exclusively localised to these boundaries.

To assess whether the epistemic uncertainty predicted by the RUGI-Unc model corresponded to regions of high registration error, we compared the highest-uncertainty and highest-error regions within each image. Substantial spatial correspondence was found in both datasets. Median exact overlap between the top 20\% most uncertain pixels and top 20\% highest-error pixels was 51.1\% for ACDC and 60.7\% for CAMUS, compared with an expected overlap of 20\% under random spatial correspondence. This corresponds to spatial enrichments of 2.55-, 3.03-fold above chance, respectively. Exact overlap was significantly greater than chance for both datasets after Benjamini-Hochberg correction. 

We additionally assessed overlap while allowing for small spatial offsets at the boundaries of the masks. Overlap increased consistently with increasing spatial tolerance across all datasets. At a tolerance of 1 pixel, median overlap increased to 69.1\% for ACDC, 76.0\% for CAMUS. At 3 pixels, overlap reached 79.7\% and 88.5\%, respectively. The rapid increase in overlap with relatively small spatial tolerances indicates that a substantial component of the incomplete exact pixelwise overlap arose from slightly displaced uncertainty and residual-error regions.

\subsubsection{Diagnostic parameters}

Table~\ref{tab:cardiac_results} shows the functional cardiac motion parameters derived from the registration deformation fields across ACDC and CAMUS.

\begin{figure}[H]
    \centering
    \includegraphics[width=0.8\linewidth]{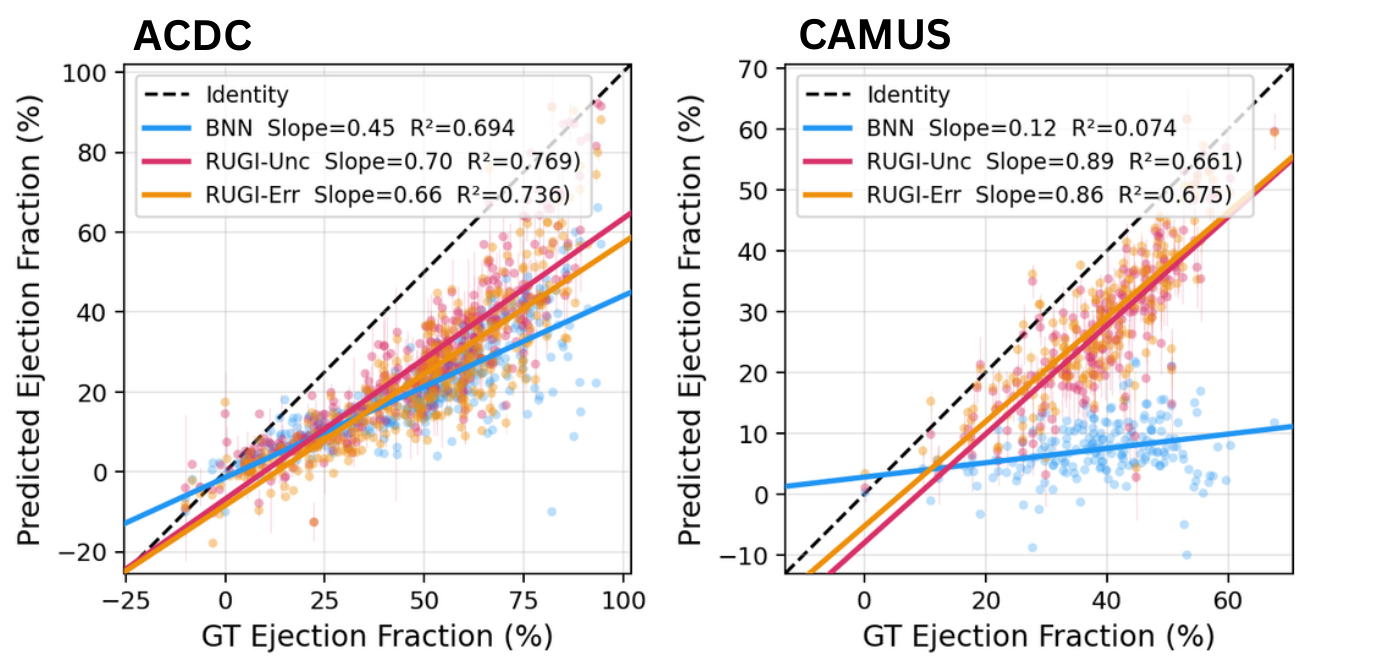}
    \caption[Ejection fraction calculated using baseline BNN and RUGI]{\textbf{Ejection fraction calculated using baseline BNN and RUGI.} Scatter plots compare EF estimates obtained from ground-truth masks compared with those obtained using the baseline BNN (blue), RUGI-Unc (green) and RUGI-Err (orange). The dashed black line represents perfect agreement. Regression lines are shown per method. }
    \label{fig:ef_scatterplots}
\end{figure}

\paragraph{Ejection Fraction.} Both RUGI variants achieved statistically significant improvements in EF estimation over the BNN baseline across both datasets. All methods systematically underestimate EF, as the more common failure mode is to not capture the full transformation. In ACDC, RUGI-Unc reduced EF MAE from 27.5\% to 14.9\%, a 45.8\% relative reduction, while RUGI-Err achieved 16.0\% MAE (Cohen's $d = 1.418$ and $1.535$ respectively, $p < 0.001$). Similar improvements were observed in CAMUS, where RUGI-Unc and RUGI-Err reduced MAE from 30.9\% to 23.0\% and 25.2\% respectively (Cohen's $d = 2.047$ for RUGI-Unc and $d = 1.522$ for RUGI-Err). All pairwise comparisons against the BNN baseline were statistically significant ($p < 0.001$, Wilcoxon signed-rank test), with effect sizes consistently in the large range ($d > 0.8$). Figure \ref{fig:ef_scatterplots} shows the predicted and ground-truth EF for all three methods. The slope of linear regression for the baseline BNN is shallow in both datasets (ACDC: $slope = 0.45$, CAMUS: $slope = 0.12$), showing the baseline model consistently underestimates the ejection fraction and it fails to capture the variance across patients. Both proposed RUGI methods achieve regression slopes closest to the identity line, with higher slopes and $R^2$ values (ACDC: $slope = 0.70$/$0.66$, CAMUS: $slope = 0.89$/$0.86$, for RUGI-Unc and RUGI-Err respectively). This shows RUGI recovers not only a lower absolute error but it also captures inter-patient variation in cardiac function. Importantly, because RUGI produces per-pixel uncertainty estimates, EF predictions can be accompanied by uncertainty intervals, which enable interpretation of the clinical parameters. 


\paragraph{Field Regularity.} The Jacobian determinant $(\bar{J})$ shows the degree of expansion and compression in the deformation field. All methods produce values close to unity across all datasets, meaning the deformations are area-preserving. RUGI variants exhibit marginally larger deviations from unity than the BNN baseline, which is expected given that iterative refinement accumulates small perturbations over multiple steps. Importantly, the percentage of voxels with negative Jacobian determinant, indicating physically impossible folding, remains below 2\% for all methods and datasets, indicating that the rate of folding remained low despite iterative refinement.

These results demonstrate that RUGI improves not only registration accuracy, as measured by EF and DSC, but also the clinical utility of the output by providing per-sample uncertainty quantification. The availability of uncertainty intervals on EF and field regularity metrics represents a qualitative advance over deterministic registration methods: rather than producing a single point estimate with no indication of reliability, RUGI provides information for downstream algorithms to reason about the trustworthiness of each individual measurement.

\begin{table}[H]
\centering
\footnotesize
\setlength{\tabcolsep}{12pt}

\begin{tabular}{@{}lccc@{}}
\toprule
\textbf{Metric} &
\textbf{BNN} &
\textbf{RUGI-Unc} &
\textbf{RUGI-Err} \\
\midrule

\multicolumn{4}{c}{\textbf{ACDC}} \\
\addlinespace[2pt]
\midrule

DSC (\%) $\uparrow$
& $73.8 \pm 14.8$
& $83.7 \pm 11.3$
& $\mathbf{84.6 \pm 10.3}$ \\

EF MAE (\%)$^\dagger$ $\downarrow$
& $27.5 \pm 14.7$
& $\mathbf{14.9 \pm 12.6}$
& $16.0 \pm 11.9$ \\

EF 95\% UI (\%)
& ---
& $2.6 \pm 1.4$
& $1.2 \pm 0.6$ \\





$\bar{J}$
& $1.00 \pm 0.023$
& $1.02 \pm 0.100$
& $0.989 \pm 0.054$ \\

95\% UI of $\bar{J}$
& ---
& $2.504 \pm 2.209$
& $0.060 \pm 0.046$ \\

\midrule
\multicolumn{4}{c}{\textbf{CAMUS}} \\
\addlinespace[2pt]
\midrule

DSC (\%) $\uparrow$
& $78.2 \pm 7.5$
& $\mathbf{82.3 \pm 7.3}$
& $82.0 \pm 7.2$ \\

EF MAE (\%)$^\dagger$ $\downarrow$
& $30.9 \pm 10.7$
& $\mathbf{23.0 \pm 11.1}$
& $25.2 \pm 10.9$ \\

EF 95\% UI (\%)
& ---
& $1.7 \pm 0.5$
& $0.9 \pm 0.3$ \\





$\bar{J}$
& $0.999 \pm 0.018$
& $1.004 \pm 0.040$
& $1.002 \pm 0.032$ \\

95\% UI of $\bar{J}$
& ---
& $1.187 \pm 1.032$
& $0.032 \pm 0.015$ \\

\bottomrule
\end{tabular}

\caption[Cardiac motion parameters]{%
    \textbf{Cardiac motion parameters extracted from registration-derived
    deformation fields.}
    \textbf{Bold}: best result per metric and dataset block.
    Population std ($\pm\sigma$) reflects inter-patient variability.
    \textbf{95\,\%\,UI} is a per-sample uncertainty estimate
    available exclusively for RUGI variants (Monte Carlo field perturbation,
    Section~\ref{sec:cimethod}).
    \emph{Units:}
    DSC in \%;
    EF\,MAE in \%;
    $\bar{J}$: Jacobian determinant
    ($1$ = perfectly area-preserving).
    $^\dagger$~Pixel-count EF from 2-D segmentation masks.%
}
\label{tab:cardiac_results}
\end{table}

\section{Discussion}

Learning-based registration models are quick at inference-time, but they generally estimate the full deformation in a single forward pass and have no ability to adapt. We introduced RUGI, a framework to iteratively refine predicted deformations while modulating each update. Across cardiac MRI and echocardiography, both uncertainty- and error-gated variants consistently improved registration performance relative to single-step inference. By weighting the incremental update by the normalised uncertainty maps, RUGI concentrates refinement effort where the initial single-step prediction is least reliable, while suppressing potentially destabilizing updates in already well-registered regions. This allows RUGI to correct the residual misalignment that single-step inference obtains.

RUGI methods produced statistically significant improvements over single-step inference across an MRI and an ultrasound dataset according to all four image similarity metrics. Improvements across both studied datasets show RUGI is not restricted to a single modality, but captures a general weakness of single-step feed-forward registration: the inability to correct residual misalignment after a single forward pass. This improvement in performance is maintained when applying our error-based gating strategy to iterate on pre-trained models. 

The ablation study isolates the gating signal as a critical component driving performance gains. Replacing the learnt signal with a random map (A2) consistently produced the largest degradation, but all ablations performed worse than both the uncertainty-based and error-based methods. However, it is worth noting that even the weakest ablations, the no-gating (A1) and intensity-based (A3), still improved over the single-step baseline, demonstrating that iterative refinement itself provides a benefit independent of the gating strategy. This suggests the contributions of our method are two-fold: the iterative mechanism provides a baseline improvement, and the selective gating function further improves the registration accuracy.  

The two proposed gating strategies performed similarly overall, with neither consistently dominating according to image similarity. The uncertainty-based method is consistently better in ejection fraction estimation, but with only two datasets conclusions cannot be drawn. The uncertainty-based method requires an additional trained network but provides a gating signal specific to the registration model. The error-based method is simpler and model-agnostic, enabling its application to pretrained models. The latter is desirable when retraining is unavailable. 

Several limitations should be noted. First, the uncertainty-based RUGI requires supervised training using samples from the registration model, which adds complexity to the training pipeline and ties the uncertainty estimates to the quality of the base model. Error-based RUGI avoids this requirement but is more dependent on the image residual providing an informative signal. Second, the iterative inference increases computational cost relative to a single forward pass, which may be prohibitive in real-time applications. Third, our experiments used 2D registration; the extension to 3D (demonstrated only for the pretrained model experiments) may introduce additional computational constraints. Fourth, performance was not uniform across all architecture-metric combinations, as demonstrated in SSIM for CycleMorph. The stopping criterion is based on image similarity improvement, which may not always align with anatomical registration accuracy; incorporating landmark-based or segmentation-based stopping criteria could improve reliability in settings where ground truth deformation is partially available. 

The larger improvements observed in ACDC relative to CAMUS highlight an important limitation of the current method. Ultrasound registration is generally more challenging because of lower signal-to-noise ratio, spatially varying speckle and intensity, and out-of-plane motion that can make local correspondences ambiguous. Although RUGI still improved upon the single-step baseline, the smaller gains suggest the current strategy is less effective under these conditions. Further improvements could improve correspondence estimation in difficult regions, for example multi-scale or pyramidal refinement, further regularisation of the loss at inference, or the incorporation of structural priors such as anatomical landmarks. 

\section*{Acknowledgements}
We gratefully acknowledge John Bonnici's work on a previous version of this approach. This work was supported by the British Heart Foundation (grant 'FS/4yPhD/F/22/34178).

\section{Conclusion}

We proposed RUGI, an algorithm for uncertainty-gated deformation field refinement at inference-time. By focusing updates on regions of the images with high uncertainty, our method is able to adapt to spatially heterogeneous difficulty.  The resulting uncertainty maps provide spatially interpretable uncertainty estimates that identify spatial variation in registration reliability and can be propagated to downstream tasks. Statistically significant improvements were demonstrated across cardiac MRI and ultrasound datasets, with improvements in intensity-based metrics and anatomical alignment. The framework generalises to existing pretrained architectures without retraining, yielding MSE reductions of 27–37\% across VoxelMorph, TransMorph, and CycleMorph. The improvement in estimation of diagnostic parameters demonstrates that these differences in performance can be meaningful for downstream functional parameter estimation. This work demonstrates the value of uncertainty estimation not only to improve interpretability but also to drive performance in medical imaging tasks. 

\bibliography{references.bib}

\end{document}